\def\year{2022}\relax
\documentclass[letterpaper]{article} 
\usepackage{aaai22}  
\usepackage{times}  
\usepackage{helvet}  
\usepackage{courier}  
\usepackage[hyphens]{url}  
\usepackage{graphicx} 
\usepackage{natbib}  
\usepackage{caption} 
\DeclareCaptionStyle{ruled}{labelfont=normalfont,labelsep=colon,strut=off} 
\usepackage{algorithm}
\usepackage{algorithmic}

\usepackage{newfloat}
\usepackage{listings}
\floatstyle{ruled}
\newfloat{listing}{tb}{lst}{}
\floatname{listing}{Listing}
\title{Analyzing the Stance of Facebook Posts on Abortion Considering State-level Health and Social Compositions}

\author{Ana Aleksandric, Henry Isaac Anderson, Anisha Dangal, Gabriela Mustata Wilson, Shirin Nilizadeh}

\affiliations{University of Texas at Arlington, Arlington, TX, United States}

\usepackage{caption}
\usepackage{subcaption}

\usepackage{xcolor}
\newif\ifcomment
\commenttrue
\ifcomment
\newcommand{\shirin}[1]{{\bf \textcolor{purple}{Shirin: #1}}}
\newcommand{\gabriela}[1]{{\bf \textcolor{red}{Gabriela: #1}}}
\newcommand{\ana}[1]{{\bf \textcolor{blue}{Ana: #1}}}

\newcommand{\henry}[1]{{\bf \textcolor{brown}{Henry: #1}}}

\else
\newcommand{\shirin}[1]{}
\newcommand{\ana}[1]{}
\newcommand{\gabriela}[1]{}
\newcommand{\henry}[1]{}
\fi

\usepackage{bibentry}

\begin{document}

\maketitle

\begin{abstract}
Abortion remains one of the most controversial topics, especially after overturning Roe v. Wade ruling in the United States. Previous literature showed that the illegality of abortion could have serious consequences, as women might seek unsafe pregnancy terminations leading to increased maternal mortality rates and negative effects on their reproductive health. 
Therefore, the stances of the abortion-related Facebook posts were analyzed at the state level in the United States from May 4 until June 30, 2022, right after the Supreme Court’s decision was disclosed. In more detail, the pre-trained Transformer architecture-based model was fine-tuned on a manually labeled training set to obtain a stance detection model suitable for the collected dataset. 
Afterward, we employed appropriate statistical tests to examine the relationships between public opinion regarding abortion, abortion legality, political leaning, and factors measuring the overall population's health, health knowledge, and vulnerability per state. 
We found that states with a higher number of views against abortion also have higher infant and maternal mortality rates. Furthermore, the stance of social media posts per state is mostly matching with the current abortion laws in these states.
While aligned with existing literature, these findings indicate how public opinion, laws, and women’s and infants’ health are related, and interventions are required to educate and protect women, especially in vulnerable populations. 
\end{abstract}

\section{Introduction}

\begin{figure*}[t]
\centerline{\includegraphics[width=0.8\textwidth]{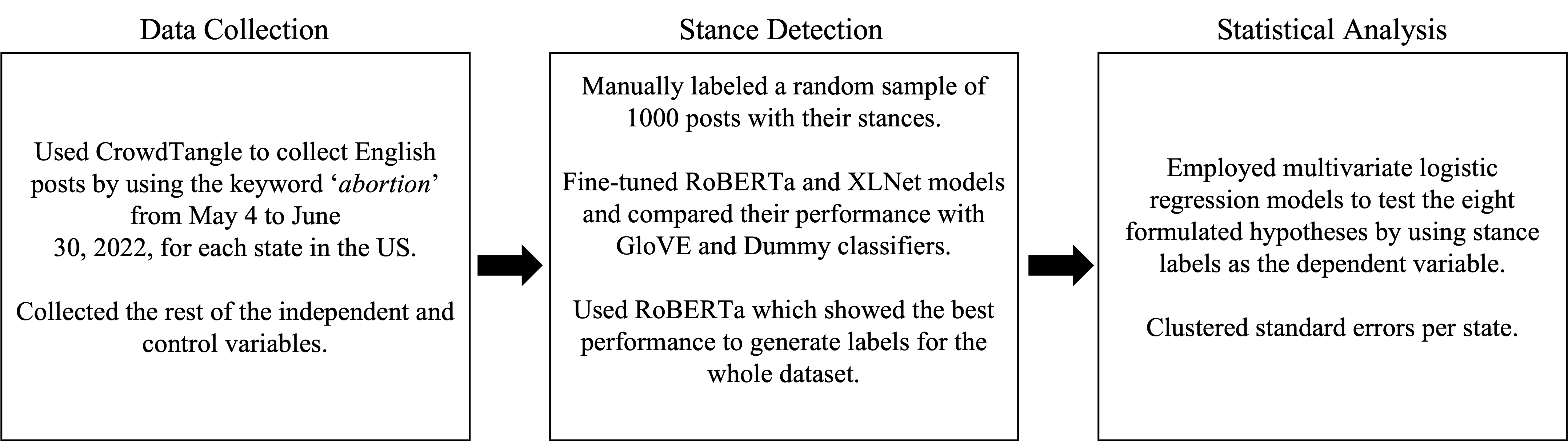}}
\caption{Study framework.}
\label{framework}
\end{figure*}

Abortion is the termination of pregnancy before the fetus reaches the viable period. Globally, about 39 abortions per thousand women take place every year and the estimation has been constant since 1990 excluding the countries that have legalized abortion which accounts for declination in 43\% of the abortion rate in those countries~\cite{council}. 
However, unsafe pregnancy terminations have been one of the leading causes of increased maternal mortality and morbidity~\cite{horga2013remarkable}. Global estimates from 2010 to 2014 demonstrate that 45\% of all abortions are unsafe abortions~\cite{who}. In addition, the restrictive abortion laws have put a financial burden on women due to which vulnerable groups of women cannot access quality care~\cite{coast2021microeconomics}, creating larger health inequities, already exacerbated by the COVID-19 pandemic. Thus, access to safe abortion is affected by numerous factors such as policies and laws related to abortion, social determinants of health (SDOH), availability of the required services, etc.~\cite{ganatra2017global}. 
Another study found that 36\% of the women lacked health insurance coverage for abortion care, and 69\% had to pay out of pocket for the care they received, making financial assistance crucial for abortion services, especially among the low-income women~\cite{jones2013cost}. 

Maternal mortality rate (MMR) is a relevant measure for the overall health of the population.
Literature shows that black women lack access to the health care services and information required to improve their reproductive health which leads to an increased Infant Mortality Rate (IMR) and Perinatal Mortality Rate~\cite{force2020maternal}.
With Roe V. Wade reversed, the MMR is expected to increase~\cite{compton2022overturning,stevenson2021pregnancy}.

Moreover, public attitudes toward abortion are related to a variety of factors, including religion, gender, stigma, political affiliation, and socioeconomic status~\cite{mosley2020attitudes,patev2019interacting}. 
Recently, some questionnaires and user studies examined attitudes toward Roe v. Wade~\cite{solon2022knowledge,crawford2022examining}. They found that more participants support Roe v. Wade than oppose it, while greater knowledge about Roe v. Wade was correlated with larger support for maintaining it.
In the literature, social media has also been shown as an effective tool to monitor public opinions on certain topics~\cite{karamouzas2022public,aldayel2021stance,chang2023roeoverturned}. 
Therefore, this study analyzes the stance of social media posts towards abortion at the state level in the US, right after the Supreme Court's draft about overturning Roe v. Wade had been disclosed. 
Afterward, we employed appropriate statistical tests to examine the relationships between public opinion regarding abortion, abortion legality, political leaning, and factors measuring the overall population's health, health knowledge, and vulnerability per state. 

Based on the literature on abortion, which will be discussed in Section Hypothesis Formulation, we define and examine the following hypotheses: 

\begin{itemize}
\item \textbf{H1:} States with a higher maternal mortality rate express less supportive stances toward abortion.
\item \textbf{H2:} States with a higher infant mortality rate express less supportive stances toward abortion.
\item \textbf{H3:} States that are mostly Republican express less supportive stances towards abortion.
\item \textbf{H4:} States with higher rape rates express more supportive stances toward abortion.
\item \textbf{H5:} States with higher social vulnerability index express less supportive stances toward abortion.
\item \textbf{H6:} States with lower health literacy express less supportive stances toward abortion.
\item \textbf{H7:} States with a lower number of abortions express less supportive stances toward abortion.
\item \textbf{H8:} States where abortion is currently illegal express less supportive stances toward abortion.
\end{itemize}

As shown in Figure~\ref{framework}, we developed a framework consisting of three modules. Firstly, data were collected from one of the most widely used social media platforms, Facebook~\cite{facebook}. Secondly, a random sample of posts was manually labeled to obtain the ground truth dataset. Thirdly, a pre-trained Transformer architecture RoBERTa model~\cite{liu2019roberta} was fine-tuned using the ground truth to create a well-performing stance detection model suitable for the collected dataset. Finally, regression analysis was employed to understand relationships between different factors and the stance of the posts at the state level in the US. Study findings imply that infant mortality rate, maternal mortality rate, political affiliation, number of abortions, and abortion legality are significantly associated with attitudes toward abortion. 
Such methodology shows how natural language processing, machine learning, and social media data can be utilized to analyze public opinion around controversial topics which can promote targeted interventions in areas where they are needed.

\section{Related Work}
\label{related}

\textbf{Stance detection models.}
Stance is defined as an individual’s standpoint toward a particular topic~\cite{aldayel2021stance}. 
Detecting stance from their social media posts is a work in progress, and many aspects of it are still unclear~\cite{aldayel2021stance}. In more detail, stance detection refers to a process of inferring the standpoint of the writer from their text by using different features~\cite{aldayel2021stance}. 
A previous study discussed the orthogonal association between stance and sentiment suggesting that e.g., positive sentiment found in the text does not necessarily imply a supportive stance towards the topic discussed~\cite{aldayel2021stance}. 
As social media are common places to share viewpoints, a lot of research focuses on developing stance detection models to understand the standpoints of users about controversial topics, such as US elections~\cite{darwish2017trump,sobhani2017dataset,lai2017friends}, and abortion~\cite{mohammad-etal-2016-semeval,stab-etal-2018-cross}. The performance of stance detection models employing supervised learning in recent studies slightly varies. 
Some approaches employing traditional machine learning models such as Support Vector Machine achieved F1 scores of 69\%~\cite{mohammad2017stance} and 63.6\%~\cite{elfardy2016cu}, while the work leveraging a bidirectional LSTM with a fast-text embedding layer reported an F1 score of 72.1\%~\cite{siddiqua2019tweet}.
However, recently, several studies approach stance detection by using BERT-based models~\cite{kawintiranon-singh-2021-knowledge,liu2021enhancing,alturayeif-etal-2022-mawqif,glandt2021stance,clark2021integrating,barbieri-etal-2020-tweeteval} reporting average F1 scores in range approximately between 0.7 to 0.9. 
Prior literature found that BERT-based models outperform other models in stance detection tasks on SemEval 2016 dataset~\cite{ghosh2019stance} and they reach state-of-the-art performance with accuracies close to or above 0.9~\cite{slovikovskaya2019transfer,dulhanty2019taking,liu2022politics}.

\textbf{Analysis of abortion online discussions}.
Previous research showed that social media might be a great tool to analyze public attitudes on different topics~\cite{karamouzas2022public,aldayel2021stance,chang2023roeoverturned}. 
For example, one study found that there was an increased interest in posting abortion-related tweets in the period of early May of 2023, before the official overturning of Roe v. Wade~\cite{mane2022examination}. Another study investigated emotions around controversial topics on online debate forums~\cite{li2020emotions}. Their findings suggest that abortion discussions contained the highest number of comments expressing an emotion of disgust compared to other emotions. In addition, previous research found that gender and political affiliation were associated with the use of incivility and intolerance in abortion referendum Twitter discussions~\cite{oh2021unpacking}. Finally, there are certain publicly available datasets that include abortion-related posts~\cite{mohammad-etal-2016-semeval,stab-etal-2018-cross,chang2023roeoverturned}.
However, to the best of our knowledge, none of the studies investigated the public stance toward abortion and state-level factors associated with it.

\section{Data Collection}
\label{data-collection}

People tend to discuss matters such as political issues and abortion on social media~\cite{karamouzas2022public,aldayel2021stance,chang2023roeoverturned,li2020emotions}. Furthermore, Facebook still remains one of the most utilized platforms~\cite{pew-research-fb}.
Therefore, the data have been collected from Facebook by using CrowdTangle~\cite{crowdtangle}, a social media insights tool that provides data from highly influential public pages, groups, and verified users. Note that CrowdTangle does not allow collecting data from personal and private accounts, or posts visible only to specific users. However, CrowdTangle lets users search the posts based on filters such as local relevance, time frame, language, keywords, etc. It is important to note that CrowdTangle finds the locations of pages/groups based on the geographic distribution of their followers on Facebook. Therefore, the English posts that contained the keyword ‘abortion’ were collected for each state in the US, from May 4 to June 30, 2022, after the leak of the Supreme Court’s decision to overturn Roe V. Wade. The total number of posts collected was 82,056.

\begin{figure}[t]
\centerline{\includegraphics[width=0.98\columnwidth]{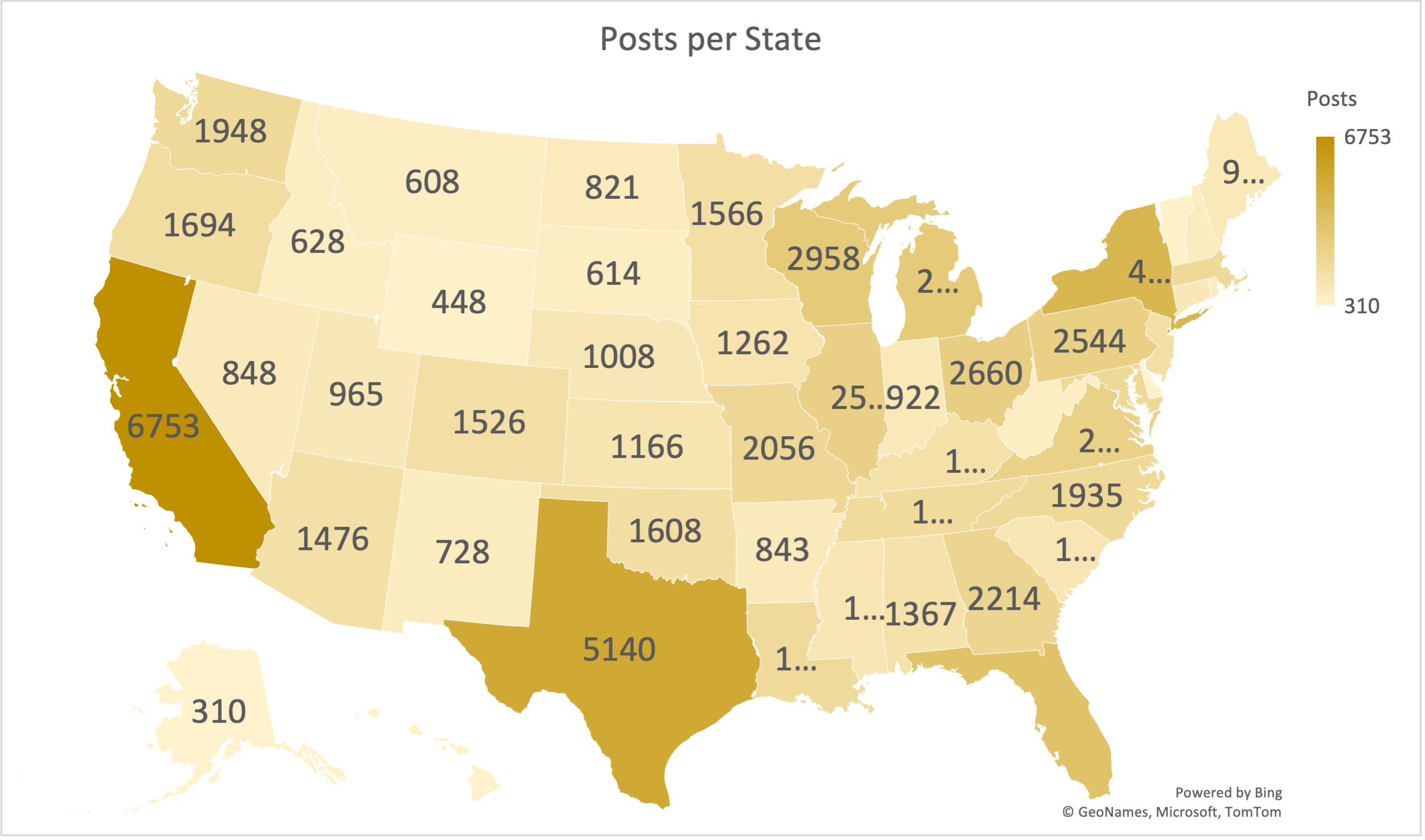}}
\caption{Number of posts per state.}
\label{posts}
\end{figure}

As indicated in Figure~\ref{posts}, certain states contain more Facebook posts collected compared to others. For example, the number of posts originating from California is the highest. 
Another state with a high number of posts is Texas, the second most populated state in the US, immediately after California. Therefore, the state population might play an important role in the variation of the number of posts per state. Therefore, it is used as a control variable in the analysis.

\section{Stance Detection}

To detect the stance of Facebook posts about abortion, we developed a model using a transfer learning approach, using XLNet~\cite{yang2019xlnet} and RoBERTa~\cite{liu2019roberta} and fine-tuning them on a ground truth dataset. While deep learning models have been used for stance detection on other topics, such as political debates~\cite{kawintiranon-singh-2021-knowledge,liu2022politics}, 
 or COVID-19 issues~\cite{glandt2021stance}, to the best of our knowledge, this work is the first developing an abortion stance detection model. 

\textbf{Stance detection models:} 
As discussed in the Related Work Section, the previous literature shows that BERT-based models achieve current state-of-the-art performance.
Thus, XLNet~\cite{yang2019xlnet} (specifically, the \emph{xlnet-base-cased} variant) and RoBERTa~\cite{liu2019roberta} (specifically, the \emph{roberta-base} variant) models were fine-tuned on the 1,000 labeled posts to create the stance detection model. 
Both are large language models based on the Transformer architecture~\cite{vaswani2017attention}, and both models regularly achieve very high accuracy on many prediction tasks. While there are studies employing XLNet~\cite{10.1145/3457682.3457717,yang2019xlnet}, previous research suggests that RoBERTa can achieve state-of-the-art results outperforming BERT and XLNet~\cite{slovikovskaya2019transfer,dulhanty2019taking,liu2022politics,barbieri-etal-2020-tweeteval}.
RoBERTa, despite its strong performance, has a maximum input length when processing texts. However, the number of posts longer than the RoBERTa's maximum length in the dataset was only 1.9\%, thus, making using RoBERTa still possible.
XLNet has no such limitation and was included as a candidate model due to the potentially uncapped length of Facebook posts.

\textbf{Groundtruth creation:} Firstly, a random sample of 1000 posts was extracted and manually labeled by three coders. The data was labeled by assigning one of the following labels to each post: stance supporting abortion, stance against abortion, and no stance. After the manual labeling by annotators was completed, the final labels for each post were determined if at least two coders assigned the same label to the same post. However, in 47 posts, labels assigned by three coders were different. Thus, annotators discussed among themselves and agreed on the final labels for such posts. In addition, we calculated a Fleiss’ Kappa~\cite{fleiss2013statistical} to better understand the inter-rater agreement. The obtained value equals 0.424 representing moderate agreement~\cite{fleiss2013statistical}. This implies that abortion stance detection is a hard task even for humans; making implementation of a well-performing stance detection model even harder.
Afterward,  we used this ground truth dataset to train and test stance detection models. 
The resulting dataset was heavily imbalanced in favor of ``no stance'' (N=470) and ``supporting'' (N=431); ``against'' stances only accounted for 99 of the messages.  The next section discusses the techniques used to mitigate the effects of this imbalance.

The dataset was split into a training, testing, and validation fold, using 80\% of the data for training, 10\% for validation, and 10\% for testing. The folds were selected at random and stratified on the stance labels. Both models were fine-tuned using the AdamW optimizer with a learning rate of $10^{-5}$, and a training batch size of 4. Early stopping was used to terminate training once the macro F1 score on the validation dataset failed to decrease for 5 training epochs, at which point the best-performing model weights were restored and evaluated on the test set.  To address the issue of data imbalance, a weighted cross-entropy loss was used \cite{aurelio2019learning}.  Cross-entropy is calculated as normal, but the per-class loss is scaled by a factor of $\frac{1}{N}$, where $N$ is the total number of training observations in that class, i.e., the final loss function takes the form:

$$
\ell(y, \hat{y}) = -\sum_{i=0}^{C} \frac{1}{N_i} log \frac{exp(\hat{y}_i)}{ \sum_{j=0}^{C}\hat{y}_j} y_i 
$$

Where $\hat{y}_i$ is the model's raw (i.e. logit) prediction for class $i$, $y_i$ is the ground truth value for class $i$ (either 0 or 1), and $N_i$ is the number of training examples in class $i$.

Due to the small size of the training dataset, the above training process was repeated 250 times to obtain a more robust measurement of the models' performance. Different training-validation-testing splits may result in markedly different model performances depending on which observations end up in which split. Repeating the training procedure multiple times is intended to measure the empirical distribution of model performance under different possible splits. After the 250 training rounds were completed, the average stopping epoch (i.e., the training epoch where the model obtained the best performance on the validation sets) was calculated and used to re-train the final model on the entire training dataset. 
Table~\ref{table:training results} shows  the summary of model performances, averaged across the 250 training runs. 

\begin{figure}[t]

     \centering
     \begin{subfigure}[b]{0.49\columnwidth}
         \includegraphics[width=\columnwidth]{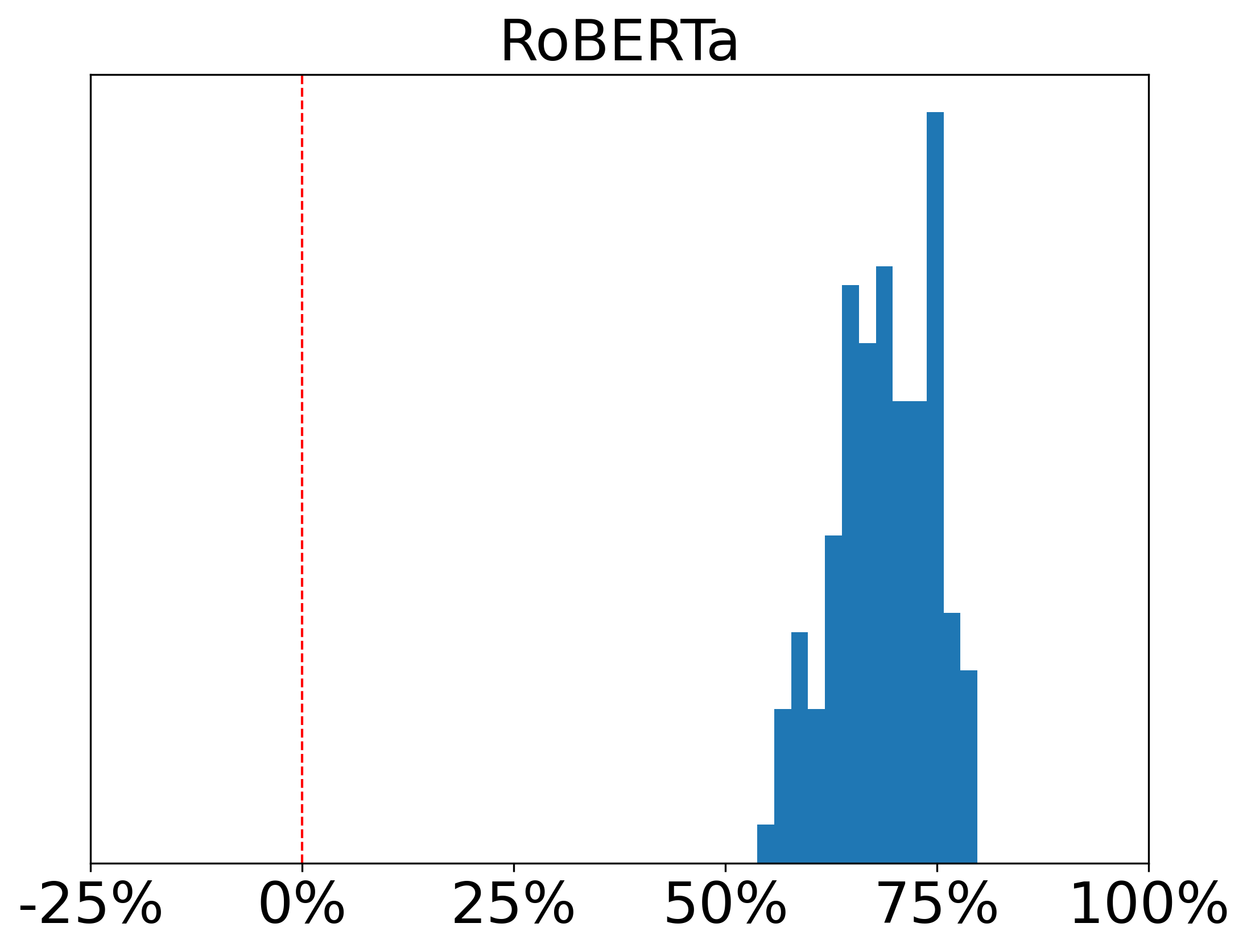}
         \caption{}
         \label{roberta absolute}
     \end{subfigure} 
         \begin{subfigure}[b]{0.49\columnwidth}
         \includegraphics[width=\columnwidth]{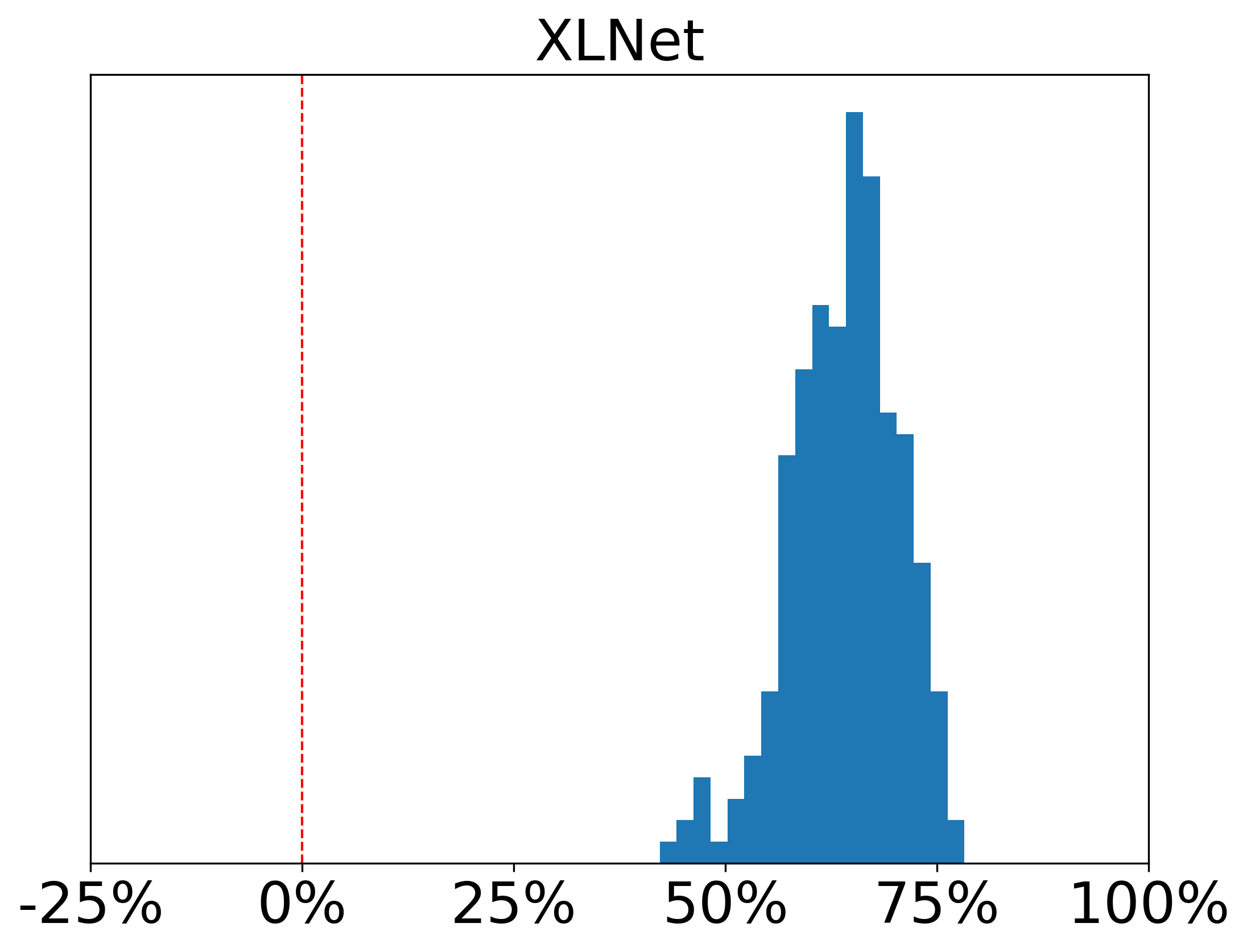}
         \caption{}
         \label{xlnet absolute}
     \end{subfigure} 

    \caption{Distributions of model scores across 250 training rounds for RoBERTa and XLNet (\ref{roberta absolute} and \ref{xlnet absolute}). A vertical reference line is at 0\%.}
    \label{figure:performance metrics}
\end{figure}


\begin{table}[h!]
    \caption{Summary of model performances across 250 random training-validation-testing splits. Standard deviations are in parentheses.}
    \label{table:training results}
    \resizebox{0.80\columnwidth}{!}{%
    \centering
    \begin{tabular}{c|cc}
      Model & Macro F1 & Balanced Accuracy  \\
         \hline
         RoBERTa & 69.01 (5.95) & 69.24 (6.43)\\
         XLNet   & 64.18 (6.50) & 63.72 (6.52) \\
    \end{tabular}
    }
\end{table}


Figure~\ref{figure:performance metrics} shows a histogram of the performance metrics' distributions. 
RoBERTa shows higher performance than XLNet, with an average F1 of 69.01 versus XLNet's 64.18, and a balanced accuracy score of 69.24 versus XLNet's 63.72. 
While these performance metrics are not competitive with the state-of-the-art stance detection models, it is important to note that the inter-rater agreement of Fleiss’ Kappa being 0.424 shows how hard it is even for three human annotators to agree on the stance of the posts. Thus, looking for a high-performance stance detection model trained on such a dataset might be an impossible expectation.

\textbf{Detecting stance of all posts:} Since the RoBERTa model obtained the highest average performance, it was selected as the final model to re-train over the entire training dataset. Following these results, a RoBERTa-base model was fine-tuned over the entire 1,000 post dataset for 5 epochs.  The fine-tuned model was then used to generate predictions for the remaining 81,056 unlabeled posts.

\begin{figure}[t]
\centerline{\includegraphics[width=0.95\columnwidth]{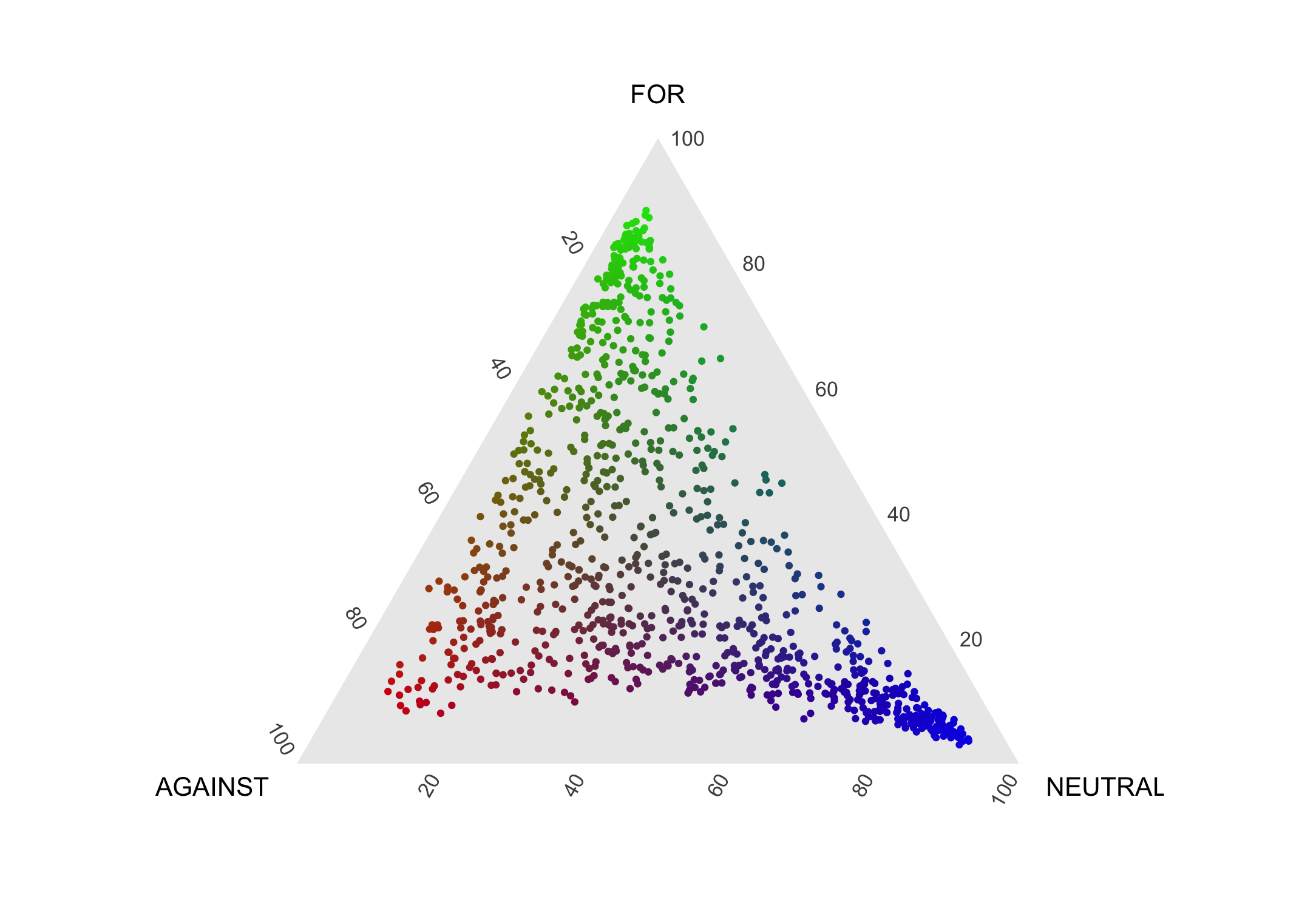}}
\caption{Scores of a random sample of posts.}
\label{scores}
\end{figure}

The following attributes were obtained for each post in the dataset: LABEL\_FOR, LABEL\_AGAINST, and LABEL\_NO\_STANCE which are adding up to 1, where each provides a likelihood of the post containing this stance about abortion. 
To better understand the distribution of posts' stances in our dataset, a ternary graph was plotted by using a random sample of 1000 posts and their scores (Figure~\ref{scores}). 
It is clear that the lowest portion of posts is against abortion, while larger numbers of posts are supporting abortion or neither. 
Finally, to label each post with its stance, the highest value of three scores was picked. Therefore, each post had a \emph{stance} of being for, against, or neutral towards abortion. The number of posts being pro-life is 22,097, while the number of posts being pro-choice is 27,493. We found 32,466 posts that did not express a stance toward abortion. 

\section{Hypothesis Formulation}
\label{hypotheses}
\begin{figure}[t!]

     \centering
     \begin{subfigure}[b]{0.47\textwidth}
         \includegraphics[width=\textwidth]{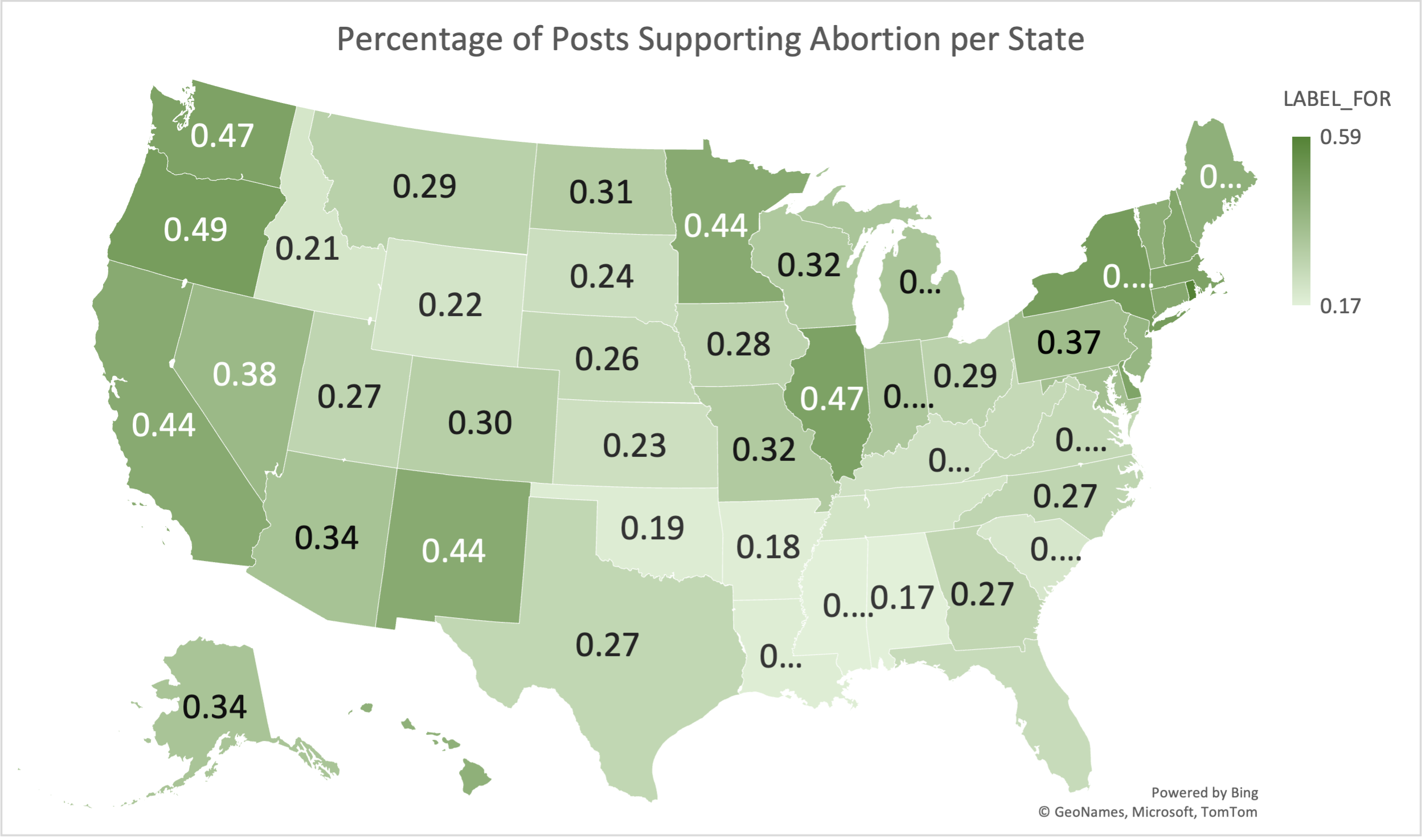}
         \caption{}
         \label{for}
     \end{subfigure} 
     \begin{subfigure}[b]{0.47\textwidth}
         \includegraphics[width=\textwidth]{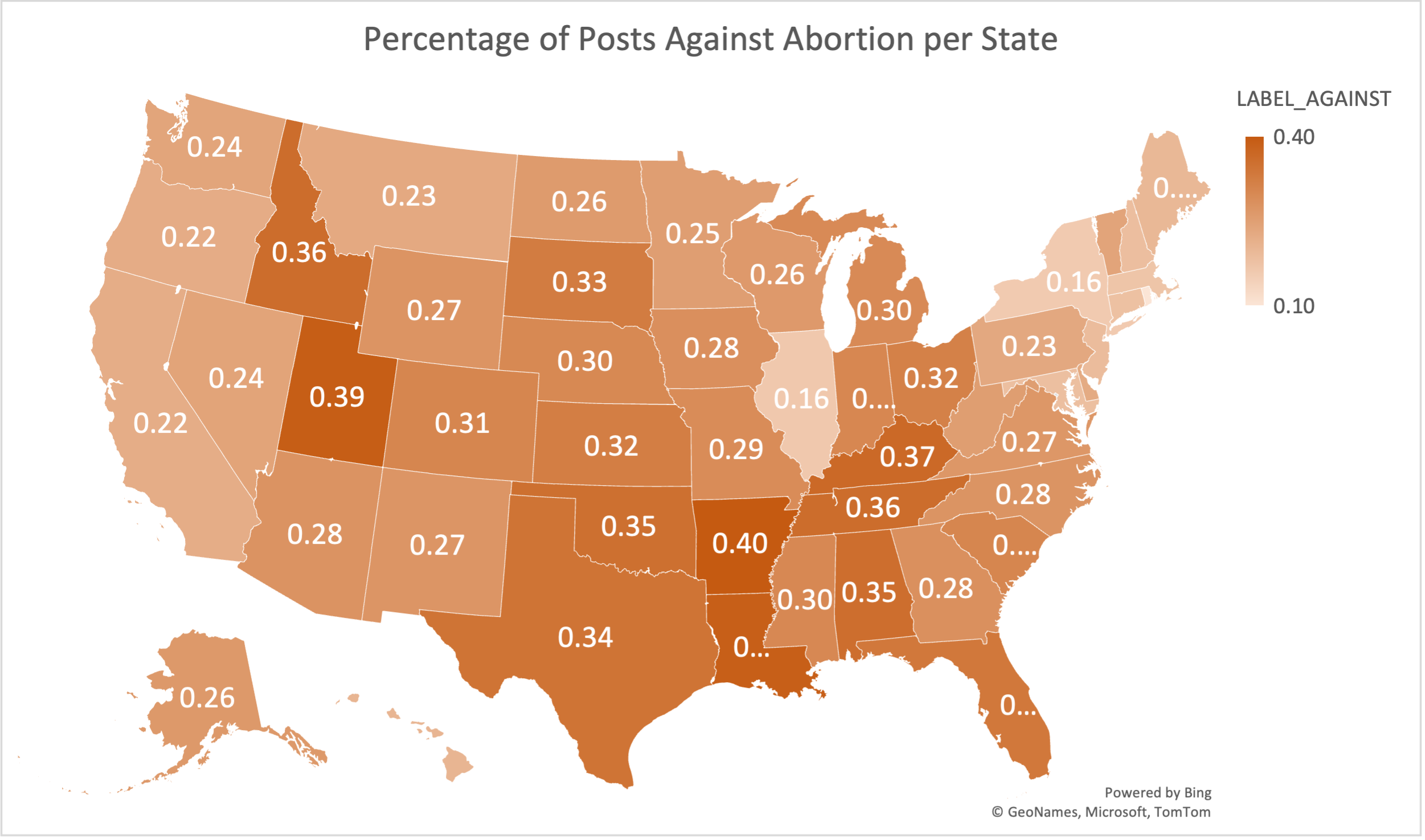}
         \caption{}
         \label{against}
     \end{subfigure}
     
    \caption{The distribution of abortion supporting and opposing stances at the state level in the US.}
\end{figure}

In this study, we use stance towards abortion as the dependent variable, as it aims to closely investigate relationships among SDOH and other factors that might affect abortion rates at the state level in the US and the stance expressed in Facebook posts originating from those states. 
The distributions of supporting and opposing stances in US states are shown in Figures~\ref{for} and~\ref{against}. Note that posts that express no stance are discarded from the analysis. 
As indicated, the distribution of stances per state is quite different. For example, the state with the highest percentage of posts against abortion is Arkansas, while the state with the highest percentage of posts supporting abortion is Rhode Island, followed by New York. 
Therefore, it is crucial to understand such differences in attitudes toward abortion among US states.

\textbf{Maternal Mortality Rate (MMR)} refers to the number of maternal deaths per 100,000 live births~\cite{mmr}. 
A study by~\citeauthor{addante2021association} found that states with more abortion restrictions had higher MMR, even after controlling for the factors such as race, income, and healthcare access. Specifically, the MMR was increased by 10\% for every additional abortion restriction in a state.
Therefore, greater economic and social gaps may be reflected in increased MMR, which may also have an impact on views toward abortion as well as accessibility to reproductive healthcare. 
Therefore, we defined the first hypothesis and used MMR as an independent variable. 
\emph{H1: States with a higher MMR express less supportive stances toward abortion}
We extracted the MMR in the US from 2018 to 2020 from the CDC~\cite{mmr-data}. Note this study employs the numeric variable \emph{Deaths} from this dataset without missing values.

\textbf{Infant Mortality Rate (IMR)} is the number of deaths of infants under the age of one per 1,000 live births in a given population~\cite{imr}. Newborns who were given birth in states with stricter abortion laws had a higher risk of infant death than those who were given birth in states with no such restrictions~\cite{pabayo2020laws}. 
Also, prior studies noted a decline in the IMR over 3 years in certain states after the legalization of abortion in 1970 in three out of five income classes~\cite{doi:10.2105/AJPH.2014.302401}.
Therefore, based on the literature, we defined the second hypothesis and used IMR as a numeric independent variable: \emph{H2: States with a higher IMR express less supportive stances toward abortion}. 
We obtained IMR rates in 2020 from Statista Research Department~\cite{imr-data}.


\textbf{Political Affiliation}.
Prior studies showed that Republicans were more inclined to support overturning Roe v. Wade whereas the Democrats had a greater likelihood to support maintaining Roe v. Wade~\cite{crawford2022examining}. 
In addition, a report from Ipsos shows that 56\% of Republicans are seemingly against the legalization of abortion while 81\% of Democrats support it~\cite{ipsos}. Therefore, political affiliation at the state level was determined by the more prevalent political party overall, and it was used as a categorical 
control variable to test the third hypothesis: \emph{H3: States that are mostly Republican express less supportive stances towards abortion}. The data on percentages of the population belonging to each party were obtained from Pew Research Center~\cite{paw-research}. 

\textbf{Rape rate} was obtained per 100,000 inhabitants in 2020 at the state level in US~\cite{rape} and was used as an independent variable in the analysis. Existing legal hurdles in many cases restricted access to abortion services by pregnant rape victims~\cite{bhate2019denial}. A study conducted to investigate the cause associated with which women opt to undergo an abortion revealed that 1\% of the total subjects had been victims~\cite{finer2005reasons}. 
Many factors contribute to a raped woman undergoing a denial process that restricts their access to obtaining abortion services, which in turn is made even worse by the complicated legal system that is currently prevalent~\cite{lara2006challenges}. 
Thus, the fourth hypothesis is as follows: \emph{H3: States with higher rape rates express more supportive stances toward abortion}.

\textbf{Social Vulnerability Index (SVI)} considers factors like income, education, housing, access to transportation, access to healthcare, etc., and measures how vulnerable populations' health might be if experiencing external stressors~\cite{svi}. A qualitative study~\cite{ouedraogo2014social} concluded that the factors that contribute to limited access to safe abortion are a poor financial background, a lack of education, ignorance of reproductive health, cultural and religious beliefs, and the legal system. Additionally, the rate of unintended pregnancy has been decreasing but it has been seen in a higher ratio among women with lower socio-economic status~\cite{finer2016declines}. As a result, SVI barriers might have an impact on accessing reproductive healthcare as well as attitudes toward abortion. The SVI data were obtained from the CDC/ATSDR SVI database the CDC/ATSDR SVI database~\cite{svi}. However, the SVI scores provided are at the US county level. Therefore, the SVI scores in this study represent a weighted average of state counties' SVI scores given their population as weights and they were used as a numeric independent variable in the analysis.
As SVI is shown as one of the factors that contribute to limited access to safe abortion, states with higher SVI might experience lower abortion rates as well as restricted abortion laws. Thus, the following hypothesis has been developed: \emph{H5: States with higher SVI express less supportive stances toward abortion}.

\textbf{Health Literacy (HL)} refers to an individual's capability to locate, comprehend, and apply the information needed to make essential decisions concerning their health~\cite{cdc-hl}. A Texas study that looked at women's knowledge, opinions, and experiences with self-induction methods for abortion revealed that many of them had heard of them, had favorable opinions of them, and that only a small percentage of them had tried them~\cite{grossman2015knowledge}. According to the study, to decrease the prevalence of unsafe abortion practices, reliable and easily accessible information regarding safe and legal abortion services is required~\cite{grossman2015knowledge}. This shows that HL has a humongous impact on how people perceive abortion policy. Finally, HL data were gathered from The University of North Carolina at Chapel Hill~\cite{hl-data} by utilizing information from the census blocks of 2010. Note that the HL data obtained is at the census block level. Therefore, a numeric independent variable at the state level is a weighted average of provided HL scores given census group population as weights.
We hypothesize that higher HL might lead to higher awareness about reproductive healthcare and its accessibility, leading to higher abortion rates in such states. This being the case, we will test the following hypothesis: \emph{H6: States with lower health literacy express less supportive stances toward abortion}.

\textbf{Number of Abortions} is affected by the restrictive abortion laws and accessibility of abortion care whereas the per capita income, the percentage of believers in Catholicism, and the proportion of individuals raised across the state have an indirect impact on the rates of abortion~\cite{gober1997role}. The abortion rate decreased from 1.61M to 1.31M in the year 1990-2000, and around the same year, the number of abortion service providers decreased by about 38\%, these figures demonstrate that the availability of the services impacts abortion rates patterns~\cite{finer2003abortion}. 
The data for the number of abortions in the US states was retrieved from the Guttmacher Institute~\cite{guttmacher}, and it has been used as a numeric independent variable in the regression model. 
Therefore, we hypothesize that states where a higher number of abortions occur practice protective abortion laws and, thus, more supporting attitudes toward abortion: \emph{H7: States with a lower number of abortions express less supportive stances toward abortion}.

\textbf{Abortion Legality} is the legality of abortion in a particular state. Women may seek unsafe abortions in nations where abortion is controlled, which can have fatal consequences. In contrast, women may have more control over their reproductive health in nations where abortion is accessible and are less likely to suffer from adverse health effects associated with unexpected pregnancies. According to a prior study~\cite{doi:10.2105/AJPH.2012.301197}, stringent abortion laws may potentially harm the health of mothers in many ways. 
Therefore, we hypothesize that the states with protective abortion laws will experience a higher number of abortions, leading to lower MMR: \emph{H8: States where abortion is currently illegal express less supportive stances toward abortion}.

Information on state-level abortion regulations in the US was gathered from the Guttmacher Institute~\cite{guttmacher}, which offers an interactive map of the US that shows the status of abortion regulations in each state as of May 4, 2023. In this map, abortion legality comprises the following labels~\cite{guttmacher}, which were used as a categorical independent variable in the analysis: 
\emph{Most restrictive} is defined as cases when states fully ban abortion.
\emph{Very restrictive} is defined as states having numerous restrictions and early gestational age ban.
\emph{Restrictive} is defined as states having numerous restrictions and later gestational age ban.
\emph{Some restrictions/protections} is defined as states either have a couple of restrictions/protections or a combination of them.
\emph{Protective} is defined as states employing some protective policies.
\emph{Very protective} is defined as states employing most of the protective policies.
\emph{Most protective} is defined as states employing all or most of the protective policies.

\textbf{Population} of each state has been used as a control variable and the data for 2022 was obtained from US Census Bureau~\cite{population}

\section{Descriptive Statistics}

This section provides descriptive statistics on variables used in the study. Table~\ref{stats} presents the minimum, median, mean, and maximum values for continuous variables.

\begin{table}[h!] \centering
\caption{Descriptive statistic on numeric variables.} 
\label{stats} 
\resizebox{0.85\columnwidth}{!}{%
\begin{tabular}{lllll}
Statistics                 & Min    & Median  & Mean & Max\\
\hline
\# Abortions & 100  & $\sim 7.6K$ &  $\sim 18.4K$ &  $\sim 154K$ \\
MMR       & 1 &  31.5 &   45.32 &  257  \\
IMR        & 0  & 5.47 &  5.37 & 8.27  \\
Rape rate   & 14.4  &  39.85  & 43.71 & 154.8  \\
SVI    & 0.14 &  0.45 &  0.47 & 0.77 \\
HL    & 235.9   & 248 & 247.2 & 256.3  \\
Population & $\sim 581K$  & $\sim 4.6M$  & $\sim 6.7M$ & $\sim 39M$ \\
\hline
\end{tabular}
}
\end{table}

The variables presented in Table~\ref{stats} highly vary by state. For example, the minimum MMR of 1 is found in Vermont, while its maximum value is 257 found in Texas. Furthermore, the mean rape rate reported is 43.71, while Alaska shows a much higher rape rate than the national average (154.8). Interestingly, the highest SVI value is 0.77 for New Mexico, which is a state with the lowest HL in the US. 
Interestingly, the top six states with the highest IMR (Mississippi, Louisiana, West Virginia, Arkansas, Alabama, and South Dakota) are 6 out of 13 states with the Most Restrictive law toward abortion. There are 22 states with SVI higher than the national average. Out of these 22 states, 9, 2, 4, and 2 states are with most restrictive, very restrictive, restrictive, and with some restrictions/~protections abortion laws.

Table~\ref{stats2} shows that there are 26 predominantly Democrat and 24 mostly Republican states in our dataset. Furthermore, the number of restrictive states, very restrictive, and most restrictive toward abortion are 11, 2, and 13, respectively. On the other hand, the number of states that are protective, very protective, and most protective are 10, 5, and 1, while there are 8 states with some restrictions/~protections. To better understand the relationship between categorical state variables, we show the number of posts per state category in Table~\ref{stats2}. Interestingly, the number of posts that support abortion in Democrat states is higher compared to the number of posts against abortion (19,665 vs. 12,742). However, the number of posts opposing abortion is higher than those supporting abortion in Republican states (7,828 vs. 9,355). Moreover, the states that are most protective, very protective, protective, or employ some restrictions/protections contain  a larger number of pro-choice posts compared to pro-life posts. On the other hand, the opposite trend is discovered in restrictive and most restrictive states.

\begin{table}[h!] \centering
\caption{Number of posts per each state category. }
\label{stats2} 
\resizebox{\columnwidth}{!}{%
\begin{tabular}{llll}
 Political Affiliations & \# States & \# Supportive Posts    & \# Opposing Posts \\
\hline
Democrat States & 26 & 19,665  & 12,742 \\
Republican States & 24 & 7,828 &  9,355  \\
& & \\
Abortion Legality   & \# States &  \# Supportive Posts    & \# Opposing Posts\\
\hline
Most Protective & 1 & 823 & 378   \\
Very Protective & 5  & 6,109 &  2,762  \\
Protective & 10 & 5,920  &  2,967 \\
Some Restrictions/Protections & 8 & 3,131   & 2,274  \\
Restrictive & 11 & 5,738  & 5,917  \\
Very Restrictive & 2 & 1,105  & 1,034  \\
Most Restrictive & 13 & 4,667  & 6,765  \\
\hline
\end{tabular}
}
\end{table}

\section{Analysis and Results}

\begin{table*}[t]
\caption{Statistical models and results. The base categories for Legality and Political affiliation are Most Protective and Democrat, respectively.}
\begin{center}
\resizebox{0.7\textwidth}{!}{%
\begin{tabular}{l l l l}
\hline
 & M1 (H1-H5) & M2 (H6 \& H7) & M3 (H8) \\
\hline
(Intercept)                           & $2.06 \; (0.36)^{***}$ & $-10.14 \; (4.17)$     & $0.78 \; (0.02)^{***}$  \\
MMR                                & $-0.00 \; (0.00)^{**}$ &        -   &  -    \\
IMR               & $-0.21 \; (0.06)^{**}$ &   -    &    -     \\
politicalRepublican                   & $-0.39 \; (0.11)^{**}$ &     -     &     -   \\
Rape rate                            & $-0.01 \; (0.00)$      &     -    &         -   \\
SVI                         & $-0.26 \; (0.32)$      &   -    & -   \\
Population                        & $0.00 \; (0.00)$       & $-0.00 \; (0.00)^{*}$  & $-0.00 \; (0.00)$   \\
HL                          &    -    & $0.04 \; (0.02)$       &              -   \\
\# Abortions                     &     -    & $0.00 \; (0.00)^{***}$ &         -     \\
Legality Most Restrictive      &   -  &   -   & $-1.15 \; (0.09)^{***}$ \\
Legality Very Restrictive     &   -   &    -    & $-0.71 \; (0.09)^{***}$ \\
Legality Restrictive       &    -     &     -    & $-0.81 \; (0.09)^{***}$ \\
Legality Some Restrictions/Protections &    -       &    -   & $-0.46 \; (0.16)^{*}$   \\
Legality Protective          &   -     &      -     & $-0.09 \; (0.11)$       \\
Legality Very Protective         & -     &    -    & $0.02 \; (0.16)$        \\
\hline
Log Likelihood    & $-32904.43$            & $-33361.32$  & $-32876.95$    \\
Num. obs.    & $49590$                & $49590$       & $49590$ \\
\hline
\multicolumn{4}{l}{\scriptsize{$^{***}p<0.0001$; $^{**}p<0.001$; $^{*}p<0.006$}}
\end{tabular}
 }
\label{results}
\end{center}
\end{table*}

Multivariate regression analysis was leveraged to test the formulated hypotheses. Each Facebook post contained the following information: \emph{stance} (supporting or opposing abortion), state the post is associated with, number of abortions, HL, SVI, IMR, MMR, rape rate, political affiliation, and legality of abortion in the state post is associated with. Firstly, we tested for multicollinearity by computing variance inflation factors (VIF). In case VIF is greater than 5 for some predictors, it indicates that these predictors are highly correlated and can cause issues in the models~\cite{gareth2013introduction}.
We calculated VIFs for all the predictor variables used in this study. VIFs found for HL, the number of abortions, and state abortion legality were 11.9, 23.9, and 29.5, respectively, suggesting that they should not be used in the same model as other variables. The additional test including just three mentioned variables and state population shows that they should not be used together (VIFs for HL equals 2.7, number of abortions 19.1, legality 6.2, and population 11.4). Even though there is a slight correlation between the number of abortions and the population, the population is still kept as a control variable as it does not change the result obtained by the model, while the abortion legality was used in the separate model that again controls for the population.

The analysis included three logistic regression models where standard errors were clustered per state, as there was value repetition for state-level variables. Note that posts containing the stance category LABEL\_NO\_STANCE were discarded from the analysis, reducing the number of posts from 82,056 to 49,590. Model 1 (M1) tested hypotheses H1-H5, by examining the relationship between SVI, IMR, MMR, rape rate, and political affiliation, with \emph{stance} as the dependent variable. And, model 2 (M2) investigated the association between HL and the number of abortions with the stance as the dependent variable, testing H6 and H7. Third model (M3) tested the hypothesis H8. All models included the state population as a control variable. 
Finally, due to using multiple predictors to examine the hypotheses, Bonferroni correction~\cite{bland1995multiple} was applied where an original p-value of 0.05 has been divided by the total number of independent and control variables (that being nine). Therefore, the variables that show a p-value lower than 0.006 in regression analysis will be considered statistically significant. 

\textbf{Testing H1-H5}. As indicated in Table~\ref{results}, the independent variables included in the model were SVI, IMR, MMR, rape rate, and political affiliation. Results suggest a significant correlation between IMR, MMR, and political affiliation with standpoints of posts regarding abortion ($p<0.001$). In more detail, a lower IMR and MMR per state are associated with a higher likelihood of pro-life Facebook posts in such states, supporting H1 and H2. Furthermore, states being predominantly Republican decreases the likelihood of abortion-supporting posts being shared in these states, supporting H3. However, the rape rate and SVI were not statistically significant ($p>0.006$), rejecting H4 and H5. 
It  might happen that the rape victims do not express their opinions regarding abortion publicly, or the percentage of abortions that occur due to rape crimes is not very high, yielding to this attribute not being a significant indicator of public stance toward abortion. In addition, states with elevated SVI might be more vulnerable to external stressors, but it is not the most impactful factor when it comes to abortion stances.
Interestingly, the top 10 states with the lowest IMR (Vermont, California, Massachusetts, New York, New Jersey, Minnesota, Oregon, Rhode Island, Washington, and Connecticut) are states that are predominantly Democrat, thus, they express a higher percentage of posts supporting abortion. 

\textbf{Testing H6 and H7}. As demonstrated in Table~\ref{results}, the independent variables included in the model were HL and the number of abortions. Model results show that there is a significant positive association between the number of abortions and the viewpoint of abortion-related Facebook posts in US states ($p<0.0001$). In other words, a higher number of abortions is increasing the likelihood of Facebook posts being supportive of abortion, supporting H7. Interestingly, in the top 11 states with the highest number of abortions, only 1 state is Republican (Georgia), suggesting that such states contain a larger percentage of posts being supportive of abortion. In contrast, HL did not show statistical significance ($p>0.006$), rejecting H6. It can be the case that the population being educated about abortion services and their accessibility does not impact individual opinion on whether terminating a pregnancy is right or wrong.  

\textbf{Testing H8}. As demonstrated in Table~\ref{results}, the independent variable included in the model was the legality of abortion in US states. Model results suggest that there is a significant association between the legality of abortion and the viewpoint of abortion-related Facebook posts in US states. As abortion legality is considered a categorical variable, the reference group is a class of \emph{Most Protective} abortion policies. The discoveries exhibit that all three categories of restrictive abortion regulations are statistically significant ($p<0.0001$) revealing that states with restricted abortion policies are more likely to contain abortion opposing posts compared to states with \emph{most protective} abortion policies. In other words, restrictive abortion state laws are correlated with a higher likelihood of disseminating pro-life posts in these states compared to most protective states, confirming H8. 

\section{Discussion}

This study examines how different factors are associated with public abortion attitudes in US states. The findings suggest that there is a statistically significant negative association between the infant mortality rate (IMR), the maternal mortality rate (MMR), states being Republican, and the legality of abortion with the stances of abortion-related posts. In contrast, the association between the number of abortions and the stances is significant and positive.

We believe that such findings might demonstrate a multi-directional relationship. In more detail, states where the population shares more posts against abortion experience lower abortion rates leading to these states expressing a higher IMR and MMR. 
Also, it is important to note that abortion rates are significantly lower in Republican states, and states where abortion access is limited.
In summary, opinions around abortions vary by state; however, the IMR and MMR might be elevated in states where abortion is restricted. Therefore, there is an urgent need for additional interventions to educate women in such states about their reproductive health and ways to avoid fatal health consequences despite the current abortion laws in their states. Note that understanding the relationships between the selected variables is a complex issue as there are multidirectional correlations and each variable is somewhat connected to another.

Despite promising results, this study still contains certain limitations. Firstly, the posts gathered to study the public stance regarding abortion are only posted in English and none of them originated from regular or private Facebook users. 
Secondly, fine-tuned stance detection model still needs some improvement to achieve current state-of-the-art performance. However, this is the first study implementing stance detection on abortion using Facebook data. 


\section{Ethical Considerations and Broader Impact}
The data includes publicly available posts shared by influential Facebook groups, pages, or verified users. The goal of the study is not to, in any way, identify any of the users who appeared in the dataset, adhering to ethical guidelines.

Furthermore, the results contribute to the general understanding of the relationships among public views concerning abortion, abortion legality, political affiliation, and attributes indicating the overall population's health, health knowledge, and vulnerability per state in the US. Hence, the study's relevance lies in identifying states that need mandatory interventions to help women avoid fatal consequences, as well as reducing IMR and MMR in such states.

\section{Conclusion}
In conclusion, the dataset gathered in this study is a collection of Facebook posts linked to abortion after the leak of the Supreme Court's decision to overturn Roe v. Wade. Afterward, a ground truth dataset was created to fine-tune the stance detection model that perform sufficiently on the collected dataset. Then, a multivariate regression analysis highlighted the significant relationships found among Facebook posts' stances toward abortion and other contributing factors at the state level in the US. The attributes that were found to be associated with public attitudes were IMR, MMR, number of abortions, abortion legality, and political leaning. Despite the limitations of this study, the results indicate that a pro-life attitude can potentially lead to negative health outcomes disclosing an urgent necessity for targeted interventions to expand the knowledge of vulnerable communities on reproductive healthcare and its accessibility.

\bibliography{aaai22}

\begin{thebibliography}{72}
\providecommand{\natexlab}[1]{#1}

\bibitem[{Addante et~al.(2021)Addante, Eisenberg, Valentine, Leonard, Maddox,
  and Hoofnagle}]{addante2021association}
Addante, A.~N.; Eisenberg, D.~L.; Valentine, M.~C.; Leonard, J.; Maddox, K.
  E.~J.; and Hoofnagle, M.~H. 2021.
\newblock The association between state-level abortion restrictions and
  maternal mortality in the United States, 1995-2017.
\newblock \emph{Contraception}, 104(5): 496--501.

\bibitem[{ALDayel and Magdy(2021)}]{aldayel2021stance}
ALDayel, A.; and Magdy, W. 2021.
\newblock Stance detection on social media: State of the art and trends.
\newblock \emph{Information Processing \& Management}, 58(4): 102597.

\bibitem[{Alturayeif, Luqman, and Ahmed(2022)}]{alturayeif-etal-2022-mawqif}
Alturayeif, N.~S.; Luqman, H.~A.; and Ahmed, M. A.~K. 2022.
\newblock Mawqif: A Multi-label {A}rabic Dataset for Target-specific Stance
  Detection.
\newblock In \emph{Proceedings of the The Seventh Arabic Natural Language
  Processing Workshop (WANLP)}, 174--184. Abu Dhabi, United Arab Emirates
  (Hybrid): Association for Computational Linguistics.

\bibitem[{Aurelio et~al.(2019)Aurelio, De~Almeida, de~Castro, and
  Braga}]{aurelio2019learning}
Aurelio, Y.~S.; De~Almeida, G.~M.; de~Castro, C.~L.; and Braga, A.~P. 2019.
\newblock Learning from imbalanced data sets with weighted cross-entropy
  function.
\newblock \emph{Neural processing letters}, 50: 1937--1949.

\bibitem[{Barbieri et~al.(2020)Barbieri, Camacho-Collados, Espinosa~Anke, and
  Neves}]{barbieri-etal-2020-tweeteval}
Barbieri, F.; Camacho-Collados, J.; Espinosa~Anke, L.; and Neves, L. 2020.
\newblock {T}weet{E}val: Unified Benchmark and Comparative Evaluation for Tweet
  Classification.
\newblock In \emph{Findings of the Association for Computational Linguistics:
  EMNLP 2020}, 1644--1650. Online: Association for Computational Linguistics.

\bibitem[{Bhate-Deosthali and Rege(2019)}]{bhate2019denial}
Bhate-Deosthali, P.; and Rege, S. 2019.
\newblock Denial of safe abortion to survivors of rape in India.
\newblock \emph{Health and human rights}, 21(2): 189.

\bibitem[{Bland and Altman(1995)}]{bland1995multiple}
Bland, J.~M.; and Altman, D.~G. 1995.
\newblock Multiple significance tests: the Bonferroni method.
\newblock \emph{Bmj}, 310(6973): 170.

\bibitem[{{Centers for Disease Control and
  Prevention}(2022{\natexlab{a}})}]{imr}
{Centers for Disease Control and Prevention}. 2022{\natexlab{a}}.
\newblock Infant Mortality.
\newblock
  \url{https://www.cdc.gov/reproductivehealth/maternalinfanthealth/infantmortality.htm},
  Accessed on 2023-05-05.

\bibitem[{{Centers for Disease Control and
  Prevention}(2022{\natexlab{b}})}]{mmr}
{Centers for Disease Control and Prevention}. 2022{\natexlab{b}}.
\newblock Maternal mortality rates in the United States, 2020.
\newblock
  \url{https://www.cdc.gov/nchs/data/hestat/maternal-mortality/2020/maternal-mortality-rates-2020.htm},
  Accessed on 2023-05-05.

\bibitem[{{Centers for Disease Control and
  Prevention}(2023{\natexlab{a}})}]{mmr-data}
{Centers for Disease Control and Prevention}. 2023{\natexlab{a}}.
\newblock Maternal deaths and mortality rates: Each state, the District of
  Columbia, United States, 2018‐2020.
\newblock
  \url{https://www.cdc.gov/nchs/maternal-mortality/mmr-2018-2020-state-data.pdf},
  Accessed on 2023-05-05.

\bibitem[{{Centers for Disease Control and
  Prevention}(2023{\natexlab{b}})}]{cdc-hl}
{Centers for Disease Control and Prevention}. 2023{\natexlab{b}}.
\newblock What Is Health Literacy?
\newblock \url{https://www.cdc.gov/healthliteracy/learn/index.html}, Accessed
  on 2023-05-08.

\bibitem[{{Centers for Disease Control and Prevention/ Agency for Toxic
  Substances and Disease Registry/ Geospatial Research, Analysis, and Services
  Program}(2018)}]{svi}
{Centers for Disease Control and Prevention/ Agency for Toxic Substances and
  Disease Registry/ Geospatial Research, Analysis, and Services Program}. 2018.
\newblock CDC/ATSDR Social Vulnerability Index.
\newblock \url{https://www.atsdr.cdc.gov/placeandhealth/svi/index.html},
  Accessed on 2023-05-08.

\bibitem[{Chang et~al.(2023)Chang, Rao, Zhong, Wojcieszak, and
  Lerman}]{chang2023roeoverturned}
Chang, R.-C.; Rao, A.; Zhong, Q.; Wojcieszak, M.; and Lerman, K. 2023.
\newblock \# RoeOverturned: Twitter Dataset on the Abortion Rights Controversy.
\newblock \emph{arXiv preprint arXiv:2302.01439}.

\bibitem[{Clark et~al.(2021)Clark, Conforti, Liu, Meng, Shareghi, and
  Collier}]{clark2021integrating}
Clark, T.; Conforti, C.; Liu, F.; Meng, Z.; Shareghi, E.; and Collier, N. 2021.
\newblock Integrating transformers and knowledge graphs for Twitter stance
  detection.
\newblock In \emph{Proceedings of the Seventh Workshop on Noisy User-generated
  Text (W-NUT 2021)}, 304--312.

\bibitem[{Coast et~al.(2021)Coast, Lattof, Meulen~Rodgers, Moore, and
  Poss}]{coast2021microeconomics}
Coast, E.; Lattof, S.~R.; Meulen~Rodgers, Y. v.~d.; Moore, B.; and Poss, C.
  2021.
\newblock The microeconomics of abortion: A scoping review and analysis of the
  economic consequences for abortion care-seekers.
\newblock \emph{Plos one}, 16(6): e0252005.

\bibitem[{Compton and Greer(2022)}]{compton2022overturning}
Compton, S.; and Greer, S.~L. 2022.
\newblock What overturning Roe v. Wade means for the United States.

\bibitem[{{Council on Foreign Relations}(2023)}]{council}
{Council on Foreign Relations}. 2023.
\newblock Abortion Law: Global Comparisons.
\newblock \url{https://www.cfr.org/article/abortion-law-global-comparisons},
  Accessed on 2023-05-04.

\bibitem[{Crawford et~al.(2022)Crawford, Jozkowski, Turner, and
  Lo}]{crawford2022examining}
Crawford, B.~L.; Jozkowski, K.~N.; Turner, R.~C.; and Lo, W.-J. 2022.
\newblock Examining the relationship between Roe v. Wade knowledge and
  sentiment across political party and abortion identity.
\newblock \emph{Sexuality Research and Social Policy}, 19(3): 837--848.

\bibitem[{{CrowdTangle Team}(2022)}]{crowdtangle}
{CrowdTangle Team}. 2022.
\newblock CrowdTangle. Facebook, Menlo Park, California, United States.

\bibitem[{Darwish, Magdy, and Zanouda(2017)}]{darwish2017trump}
Darwish, K.; Magdy, W.; and Zanouda, T. 2017.
\newblock Trump vs. Hillary: What went viral during the 2016 US presidential
  election.
\newblock In \emph{Social Informatics: 9th International Conference, SocInfo
  2017, Oxford, UK, September 13-15, 2017, Proceedings, Part I 9}, 143--161.
  Springer.

\bibitem[{Dulhanty et~al.(2019)Dulhanty, Deglint, Daya, and
  Wong}]{dulhanty2019taking}
Dulhanty, C.; Deglint, J.~L.; Daya, I.~B.; and Wong, A. 2019.
\newblock Taking a stance on fake news: Towards automatic disinformation
  assessment via deep bidirectional transformer language models for stance
  detection.
\newblock \emph{arXiv preprint arXiv:1911.11951}.

\bibitem[{Elfardy and Diab(2016)}]{elfardy2016cu}
Elfardy, H.; and Diab, M. 2016.
\newblock Cu-gwu perspective at semeval-2016 task 6: Ideological stance
  detection in informal text.
\newblock In \emph{Proceedings of the 10th international workshop on semantic
  evaluation (SemEval-2016)}, 434--439.

\bibitem[{{Facebook Inc.}(2023)}]{facebook}
{Facebook Inc.} 2023.
\newblock Facebook.

\bibitem[{Finer and Fine(2013)}]{doi:10.2105/AJPH.2012.301197}
Finer, L.; and Fine, J.~B. 2013.
\newblock Abortion Law Around the World: Progress and Pushback.
\newblock \emph{American Journal of Public Health}, 103(4): 585--589.
\newblock PMID: 23409915.

\bibitem[{Finer et~al.(2005)Finer, Frohwirth, Dauphinee, Singh, and
  Moore}]{finer2005reasons}
Finer, L.~B.; Frohwirth, L.~F.; Dauphinee, L.~A.; Singh, S.; and Moore, A.~M.
  2005.
\newblock Reasons US women have abortions: quantitative and qualitative
  perspectives.
\newblock \emph{Perspectives on sexual and reproductive health}, 37(3):
  110--118.

\bibitem[{Finer and Henshaw(2003)}]{finer2003abortion}
Finer, L.~B.; and Henshaw, S.~K. 2003.
\newblock Abortion incidence and services in the United States in 2000.
\newblock \emph{Perspectives on sexual and reproductive health}, 35(1): 6--15.

\bibitem[{Finer and Zolna(2016)}]{finer2016declines}
Finer, L.~B.; and Zolna, M.~R. 2016.
\newblock Declines in unintended pregnancy in the United States, 2008--2011.
\newblock \emph{New England journal of medicine}, 374(9): 843--852.

\bibitem[{Fleiss, Levin, and Paik(2013)}]{fleiss2013statistical}
Fleiss, J.~L.; Levin, B.; and Paik, M.~C. 2013.
\newblock \emph{Statistical methods for rates and proportions}.
\newblock john wiley \& sons.

\bibitem[{Ganatra et~al.(2017)Ganatra, Gerdts, Rossier, Johnson, Tun{\c{c}}alp,
  Assifi, Sedgh, Singh, Bankole, Popinchalk et~al.}]{ganatra2017global}
Ganatra, B.; Gerdts, C.; Rossier, C.; Johnson, B.~R.; Tun{\c{c}}alp, {\"O}.;
  Assifi, A.; Sedgh, G.; Singh, S.; Bankole, A.; Popinchalk, A.; et~al. 2017.
\newblock Global, regional, and subregional classification of abortions by
  safety, 2010--14: estimates from a Bayesian hierarchical model.
\newblock \emph{The Lancet}, 390(10110): 2372--2381.

\bibitem[{Gareth et~al.(2013)Gareth, Daniela, Trevor, and
  Robert}]{gareth2013introduction}
Gareth, J.; Daniela, W.; Trevor, H.; and Robert, T. 2013.
\newblock \emph{An introduction to statistical learning: with applications in
  R}.
\newblock Spinger.

\bibitem[{Ghosh et~al.(2019)Ghosh, Singhania, Singh, Rudra, and
  Ghosh}]{ghosh2019stance}
Ghosh, S.; Singhania, P.; Singh, S.; Rudra, K.; and Ghosh, S. 2019.
\newblock Stance detection in web and social media: a comparative study.
\newblock In \emph{Experimental IR Meets Multilinguality, Multimodality, and
  Interaction: 10th International Conference of the CLEF Association, CLEF
  2019, Lugano, Switzerland, September 9--12, 2019, Proceedings 10}, 75--87.
  Springer.

\bibitem[{Glandt et~al.(2021)Glandt, Khanal, Li, Caragea, and
  Caragea}]{glandt2021stance}
Glandt, K.; Khanal, S.; Li, Y.; Caragea, D.; and Caragea, C. 2021.
\newblock Stance detection in COVID-19 tweets.
\newblock In \emph{Proceedings of the 59th Annual Meeting of the Association
  for Computational Linguistics and the 11th International Joint Conference on
  Natural Language Processing (Long Papers)}, volume~1.

\bibitem[{Gober(1997)}]{gober1997role}
Gober, P. 1997.
\newblock The role of access in explaining state abortion rates.
\newblock \emph{Social Science \& Medicine}, 44(7): 1003--1016.

\bibitem[{Grossman et~al.(2015)Grossman, Hendrick, Fuentes, White, Hopkins,
  Stevenson, Lopez, Yeatman, and Potter}]{grossman2015knowledge}
Grossman, D.; Hendrick, E.; Fuentes, L.; White, K.; Hopkins, K.; Stevenson, A.;
  Lopez, C.~H.; Yeatman, S.; and Potter, J. 2015.
\newblock Knowledge, opinion and experience related to abortion self-induction
  in Texas.
\newblock \emph{Contraception}, 92(4): 360--361.

\bibitem[{{Guttmacher Institute}(2023)}]{guttmacher}
{Guttmacher Institute}. 2023.
\newblock Interactive Map: US Abortion Policies and Access After Roe.
\newblock
  \url{https://states.guttmacher.org/policies/texas/abortion-statistics},
  Accessed on 2023-05-05.

\bibitem[{Horga, Gerdts, and Potts(2013)}]{horga2013remarkable}
Horga, M.; Gerdts, C.; and Potts, M. 2013.
\newblock The remarkable story of Romanian women's struggle to manage their
  fertility.
\newblock \emph{Journal of Family Planning and Reproductive Health Care},
  39(1): 2--4.

\bibitem[{{Ipsos}(2023)}]{ipsos}
{Ipsos}. 2023.
\newblock Nearly 7 in 10 support state-level ballot measures on abortion.
\newblock
  \url{https://www.ipsos.com/en-us/nearly-7-10-support-state-level-ballot-measures-abortion},
  Accessed on 2023-05-05.

\bibitem[{Jones, Upadhyay, and Weitz(2013)}]{jones2013cost}
Jones, R.~K.; Upadhyay, U.~D.; and Weitz, T.~A. 2013.
\newblock At what cost? Payment for abortion care by US women.
\newblock \emph{Women's Health Issues}, 23(3): e173--e178.

\bibitem[{Karamouzas, Mademlis, and Pitas(2022)}]{karamouzas2022public}
Karamouzas, D.; Mademlis, I.; and Pitas, I. 2022.
\newblock Public opinion monitoring through collective semantic analysis of
  tweets.
\newblock \emph{Social Network Analysis and Mining}, 12(1): 91.

\bibitem[{Kawintiranon and Singh(2021)}]{kawintiranon-singh-2021-knowledge}
Kawintiranon, K.; and Singh, L. 2021.
\newblock Knowledge Enhanced Masked Language Model for Stance Detection.
\newblock In \emph{Proceedings of the 2021 Conference of the North American
  Chapter of the Association for Computational Linguistics: Human Language
  Technologies}, 4725--4735. Online: Association for Computational Linguistics.

\bibitem[{Krieger et~al.(2015)Krieger, Gruskin, Singh, Kiang, Chen, Waterman,
  Gottlieb, Beckfield, and Coull}]{doi:10.2105/AJPH.2014.302401}
Krieger, N.; Gruskin, S.; Singh, N.; Kiang, M.~V.; Chen, J.~T.; Waterman,
  P.~D.; Gottlieb, J.; Beckfield, J.; and Coull, B.~A. 2015.
\newblock Reproductive Justice and the Pace of Change: Socioeconomic Trends in
  US Infant Death Rates by Legal Status of Abortion, 1960–1980.
\newblock \emph{American Journal of Public Health}, 105(4): 680--682.
\newblock PMID: 25713932.

\bibitem[{Lai et~al.(2017)Lai, Hern{\'a}ndez~Far{\'\i}as, Patti, and
  Rosso}]{lai2017friends}
Lai, M.; Hern{\'a}ndez~Far{\'\i}as, D.~I.; Patti, V.; and Rosso, P. 2017.
\newblock Friends and enemies of clinton and trump: using context for detecting
  stance in political tweets.
\newblock In \emph{Advances in Computational Intelligence: 15th Mexican
  International Conference on Artificial Intelligence, MICAI 2016, Canc{\'u}n,
  Mexico, October 23--28, 2016, Proceedings, Part I 15}, 155--168. Springer.

\bibitem[{Lara et~al.(2006)Lara, Garc{\'\i}a, Ortiz, and
  Yam}]{lara2006challenges}
Lara, D.; Garc{\'\i}a, S.; Ortiz, O.; and Yam, E.~A. 2006.
\newblock Challenges accessing legal abortion after rape in Mexico City.
\newblock \emph{Gaceta M{\'e}dica de M{\'e}xico}, 142: 85--89.

\bibitem[{Li and Xiao(2020)}]{li2020emotions}
Li, J.; and Xiao, L. 2020.
\newblock Emotions in online debates: Tales from 4Forums and ConvinceMe.
\newblock \emph{Proceedings of the Association for Information Science and
  Technology}, 57(1): e255.

\bibitem[{Liu et~al.(2021)Liu, Lin, Tan, and Wang}]{liu2021enhancing}
Liu, R.; Lin, Z.; Tan, Y.; and Wang, W. 2021.
\newblock Enhancing zero-shot and few-shot stance detection with commonsense
  knowledge graph.
\newblock In \emph{Findings of the Association for Computational Linguistics:
  ACL-IJCNLP 2021}, 3152--3157.

\bibitem[{Liu et~al.(2019)Liu, Ott, Goyal, Du, Joshi, Chen, Levy, Lewis,
  Zettlemoyer, and Stoyanov}]{liu2019roberta}
Liu, Y.; Ott, M.; Goyal, N.; Du, J.; Joshi, M.; Chen, D.; Levy, O.; Lewis, M.;
  Zettlemoyer, L.; and Stoyanov, V. 2019.
\newblock Roberta: A robustly optimized bert pretraining approach.
\newblock \emph{arXiv preprint arXiv:1907.11692}.

\bibitem[{Liu et~al.(2022)Liu, Zhang, Wegsman, Beauchamp, and
  Wang}]{liu2022politics}
Liu, Y.; Zhang, X.~F.; Wegsman, D.; Beauchamp, N.; and Wang, L. 2022.
\newblock POLITICS: pretraining with same-story article comparison for ideology
  prediction and stance detection.
\newblock \emph{arXiv preprint arXiv:2205.00619}.

\bibitem[{Mane et~al.(2022)Mane, Yue, Yu, Doig, Wei, Delcid, Harris, Nguyen,
  and Nguyen}]{mane2022examination}
Mane, H.; Yue, X.; Yu, W.; Doig, A.~C.; Wei, H.; Delcid, N.; Harris, A.-G.;
  Nguyen, T.~T.; and Nguyen, Q.~C. 2022.
\newblock Examination of the Public’s Reaction on Twitter to the Over-Turning
  of Roe v Wade and Abortion Bans.
\newblock In \emph{Healthcare}, volume~10, 2390. MDPI.

\bibitem[{{Maternal Health Task Force}(2020)}]{force2020maternal}
{Maternal Health Task Force}. 2020.
\newblock Maternal health in the United States.
\newblock
  \url{https://www.mhtf.org/topics/maternal-health-in-the-united-states/},
  Accessed on 2023-05-04.

\bibitem[{Mohammad et~al.(2016)Mohammad, Kiritchenko, Sobhani, Zhu, and
  Cherry}]{mohammad-etal-2016-semeval}
Mohammad, S.; Kiritchenko, S.; Sobhani, P.; Zhu, X.; and Cherry, C. 2016.
\newblock {S}em{E}val-2016 Task 6: Detecting Stance in Tweets.
\newblock In \emph{Proceedings of the 10th International Workshop on Semantic
  Evaluation ({S}em{E}val-2016)}, 31--41. San Diego, California: Association
  for Computational Linguistics.

\bibitem[{Mohammad, Sobhani, and Kiritchenko(2017)}]{mohammad2017stance}
Mohammad, S.~M.; Sobhani, P.; and Kiritchenko, S. 2017.
\newblock Stance and sentiment in tweets.
\newblock \emph{ACM Transactions on Internet Technology (TOIT)}, 17(3): 1--23.

\bibitem[{Mosley et~al.(2020)Mosley, Anderson, Harris, Fleming, and
  Schulz}]{mosley2020attitudes}
Mosley, E.~A.; Anderson, B.~A.; Harris, L.~H.; Fleming, P.~J.; and Schulz,
  A.~J. 2020.
\newblock Attitudes toward abortion, social welfare programs, and gender roles
  in the US and South Africa.
\newblock \emph{Critical public health}, 30(4): 441--456.

\bibitem[{{National Health Literacy Mapping to Inform Health Care
  Policy}(2014)}]{hl-data}
{National Health Literacy Mapping to Inform Health Care Policy}. 2014.
\newblock Health Literacy Data Map. University of North Carolina at Chapel
  Hill.
\newblock \url{http://healthliteracymap.unc.edu/#}, Accessed on 2023-05-08.

\bibitem[{Oh et~al.(2021)Oh, Elayan, Sykora, and Downey}]{oh2021unpacking}
Oh, D.; Elayan, S.; Sykora, M.; and Downey, J. 2021.
\newblock Unpacking uncivil society: Incivility and intolerance in the 2018
  Irish abortion referendum discussions on Twitter.
\newblock \emph{Nordicom Review}, 42(s1): 103--118.

\bibitem[{Ou{\'e}draogo and Sundby(2014)}]{ouedraogo2014social}
Ou{\'e}draogo, R.; and Sundby, J. 2014.
\newblock Social determinants and access to induced abortion in Burkina Faso:
  from two case studies.
\newblock \emph{Obstetrics and gynecology international}, 2014.

\bibitem[{Pabayo et~al.(2020)Pabayo, Ehntholt, Cook, Reynolds, Muennig, and
  Liu}]{pabayo2020laws}
Pabayo, R.; Ehntholt, A.; Cook, D.~M.; Reynolds, M.; Muennig, P.; and Liu,
  S.~Y. 2020.
\newblock Laws restricting access to abortion services and infant mortality
  risk in the United States.
\newblock \emph{International Journal of Environmental Research and Public
  Health}, 17(11): 3773.

\bibitem[{Patev, Hood, and Hall(2019)}]{patev2019interacting}
Patev, A.~J.; Hood, K.~B.; and Hall, C.~J. 2019.
\newblock The interacting roles of abortion stigma and gender on attitudes
  toward abortion legality.
\newblock \emph{Personality and Individual Differences}, 146: 87--92.

\bibitem[{{Pew Research Center}(2021)}]{pew-research-fb}
{Pew Research Center}. 2021.
\newblock Social Media Use in 2021.
\newblock
  \url{https://www.pewresearch.org/internet/2021/04/07/social-media-use-in-2021/},
  Accessed on 2023-05-15.

\bibitem[{{Pew Research Center}(n.d.)}]{paw-research}
{Pew Research Center}. n.d.
\newblock Party affiliation by state.
\newblock
  \url{https://www.pewresearch.org/religion/religious-landscape-study/compare/party-affiliation/by/state/},
  Accessed on 2023-05-05.

\bibitem[{Siddiqua, Chy, and Aono(2019)}]{siddiqua2019tweet}
Siddiqua, U.~A.; Chy, A.~N.; and Aono, M. 2019.
\newblock Tweet stance detection using an attention based neural ensemble
  model.
\newblock In \emph{Proceedings of the 2019 conference of the north American
  chapter of the association for computational linguistics: Human language
  technologies, volume 1 (long and short papers)}, 1868--1873.

\bibitem[{Slovikovskaya(2019)}]{slovikovskaya2019transfer}
Slovikovskaya, V. 2019.
\newblock Transfer learning from transformers to fake news challenge stance
  detection (FNC-1) task.
\newblock \emph{arXiv preprint arXiv:1910.14353}.

\bibitem[{Sobhani, Inkpen, and Zhu(2017)}]{sobhani2017dataset}
Sobhani, P.; Inkpen, D.; and Zhu, X. 2017.
\newblock A dataset for multi-target stance detection.
\newblock In \emph{Proceedings of the 15th Conference of the European Chapter
  of the Association for Computational Linguistics: Volume 2, Short Papers},
  551--557.

\bibitem[{Solon et~al.(2022)Solon, Kaplan, Crawford, Turner, Lo, and
  Jozkowski}]{solon2022knowledge}
Solon, M.; Kaplan, A.~M.; Crawford, B.~L.; Turner, R.~C.; Lo, W.-J.; and
  Jozkowski, K.~N. 2022.
\newblock Knowledge of and Attitudes Toward Roe v. Wade Among US Latinx Adults.
\newblock \emph{Hispanic Journal of Behavioral Sciences}, 44(1): 71--93.

\bibitem[{Stab et~al.(2018)Stab, Miller, Schiller, Rai, and
  Gurevych}]{stab-etal-2018-cross}
Stab, C.; Miller, T.; Schiller, B.; Rai, P.; and Gurevych, I. 2018.
\newblock Cross-topic Argument Mining from Heterogeneous Sources.
\newblock In \emph{Proceedings of the 2018 Conference on Empirical Methods in
  Natural Language Processing}, 3664--3674. Brussels, Belgium: Association for
  Computational Linguistics.

\bibitem[{{Statista Research Department}(2022{\natexlab{a}})}]{rape}
{Statista Research Department}. 2022{\natexlab{a}}.
\newblock Forcible rape rate per 100,000 inhabitants in the United States in
  2020, by state.
\newblock
  \url{https://www.statista.com/statistics/232563/forcible-rape-rate-in-the-us-by-state/},
  Accessed on 2023-05-08.

\bibitem[{{Statista Research Department}(2022{\natexlab{b}})}]{imr-data}
{Statista Research Department}. 2022{\natexlab{b}}.
\newblock Infant mortality rate in the United States as of 2020, by state
  (deaths per 1,000 live births).
\newblock
  \url{https://www.statista.com/statistics/252064/us-infant-mortality-rate-by-ethnicity-2011/},
  Accessed on 2023-05-10.

\bibitem[{Stevenson(2021)}]{stevenson2021pregnancy}
Stevenson, A.~J. 2021.
\newblock The pregnancy-related mortality impact of a total abortion ban in the
  United States: a research note on increased deaths due to remaining pregnant.
\newblock \emph{Demography}, 58(6): 2019--2028.

\bibitem[{SU et~al.(2021)SU, XI, CAO, TANG, and PAN}]{10.1145/3457682.3457717}
SU, Z.; XI, Y.; CAO, R.; TANG, H.; and PAN, H. 2021.
\newblock A Stance Detection Approach Based on Generalized Autoregressive
  Pretrained Language Model in Chinese Microblogs.
\newblock In \emph{2021 13th International Conference on Machine Learning and
  Computing}, ICMLC 2021, 232–238. New York, NY, USA: Association for
  Computing Machinery.
\newblock ISBN 9781450389310.

\bibitem[{{United States Census Bureau}(2023)}]{population}
{United States Census Bureau}. 2023.
\newblock State Population Totals and Components of Change: 2020-2022.
\newblock
  \url{https://www.census.gov/data/tables/time-series/demo/popest/2020s-state-total.html},
  Accessed on 2023-05-08.

\bibitem[{Vaswani et~al.(2017)Vaswani, Shazeer, Parmar, Uszkoreit, Jones,
  Gomez, Kaiser, and Polosukhin}]{vaswani2017attention}
Vaswani, A.; Shazeer, N.; Parmar, N.; Uszkoreit, J.; Jones, L.; Gomez, A.~N.;
  Kaiser, {\L}.; and Polosukhin, I. 2017.
\newblock Attention is all you need.
\newblock \emph{Advances in neural information processing systems}, 30.

\bibitem[{{World Health Organization}(n.d.)}]{who}
{World Health Organization}. n.d.
\newblock Maternal mortality ratio (per 100 000 live births).
\newblock
  \url{https://www.who.int/data/gho/indicator-metadata-registry/imr-details/26},
  Accessed on 2023-05-04.

\bibitem[{Yang et~al.(2019)Yang, Dai, Yang, Carbonell, Salakhutdinov, and
  Le}]{yang2019xlnet}
Yang, Z.; Dai, Z.; Yang, Y.; Carbonell, J.; Salakhutdinov, R.~R.; and Le, Q.~V.
  2019.
\newblock Xlnet: Generalized autoregressive pretraining for language
  understanding.
\newblock \emph{Advances in neural information processing systems}, 32.

\end{thebibliography}

\end{document}